\documentclass[aps,pra,twocolumn,amsmath,amssymb]{revtex4-1}

\usepackage{amsmath, amsthm, amsfonts}
\usepackage{graphicx}
\usepackage{color,amssymb}
\usepackage{psfrag}
\usepackage{longtable}
\usepackage{dcolumn,epsfig}
\usepackage[normalem]{ulem}
\usepackage{color}

\newcommand{\be}{\begin{equation}}
\newcommand{\ee}{\end{equation}}
\newcommand{\ba}{\begin{eqnarray}}
\newcommand{\ea}{\end{eqnarray}}

\newcommand{\ban}{\begin{eqnarray*}}
\newcommand{\ean}{\end{eqnarray*}}

\begin{document}

\title{Parametric Instability in Long Optical Cavities and Suppression by Dynamic Transverse Mode Frequency Modulation}
\author{Chunnong Zhao, Li Ju, Qi Fang, Carl Blair, Jiayi Qin, David Blair}
\affiliation{School of Physics, University of Western Australia, WA 6009, Australia}
\author{Jerome Degallaix}
\affiliation{Laboratoire des Mat\'{e}riaux Avanc\'{e}s, IN2P3/CNRS, Universit\'{e} de Lyon, Villeurbanne, France}
\author{Hiroaki Yamamoto}
\affiliation{Theoretical Physics \& Simulation Group, LIGO Caltech, MC 100-36,Pasadena CA 91125, USA}

\date{December 23, 2015}

\begin{abstract}

Three mode parametric instability has been predicted in Advanced gravitational wave detectors. Here we present the first observation of this phenomenon in a large scale suspended optical cavity designed to be comparable to those of advanced gravitational wave detectors.  Our results show that previous modelling assumptions that transverse optical modes are stable in frequency except for frequency drifts on a thermal deformation time scale is unlikely to be valid for suspended mass optical cavities. We demonstrate that mirror figure errors cause a dependence of transverse mode offset frequency on spot position. Combined with low frequency residual motion of suspended mirrors, this leads to transverse mode frequency modulation which suppresses the effective parametric gain. We show that this gain suppression mechanism can be enhanced by laser spot dithering or fast thermal modulation. Using Advanced LIGO test mass data and thermal modelling we show that gain suppression factors of 10-20 could be achieved for individual modes, sufficient to greatly ameliorate the parametric instability problem.
\end{abstract}

\maketitle

\section{Introduction}

Advanced laser interferometer gravitational wave detectors are currently being commissioned \cite{aLIGO,aVirgo}. Once they reach target sensitivity they have a high probability of observing gravitational waves, especially from the coalescence of binary neutron stars. Target sensitivity requires very high optical power in the detector optical cavities, which can allow radiation pressure induced instabilities.

In 2001, Braginsky \emph{et al.}  \cite{Braginsky2001, Braginsky2002} predicted that opto-acoustic interactions in such detectors could lead to a new form of instability called a three-mode parametric instability. It could arise from optical transitions between cavity modes mediated by test mass acoustic modes. Specifically photons from the main interferometer pump mode are scattered from thermally excited acoustic modes in the test masses. The pump photon creates a phonon-photon pair. If the phonon is resonant in a test mass acoustic mode, and the photon is resonant in an interferometer cavity transverse mode, this scattering process will occur resonantly. Assuming that the scattered photons have lower frequency than the pump photons, energy conservation requires the phonon to increase the occupation number of the acoustic mode. If the acoustic energy injection by this mechanism exceeds the characteristic losses of the acoustic mode, the scattering will lead to an exponential growth of the acoustic mode occupation number. Braginsky showed that the amplitude of this scattering process could be large if the spatial 2D surface amplitude distribution of the acoustic mode overlapped the spatial intensity distribution of the optical mode, thereby causing three mode parametric instability.

Subsequently Zhao \emph{et al.} \cite{zhao2005} demonstrated that for realistic interferometer designs there was a substantial risk of instability, because the high acoustic mode density in the 50-150kHz range led to numerous accidental overlaps of both mode shape and frequency. Such instabilities could not be completely avoided through optical design. This led to research focussed on observation and study of three mode interactions \cite{zhaoPRA2008}, and on methods for suppressing instability \cite{Degallaix, Evans, GrasDamper, absorber, JuCQGPI}.

Strigin \emph{et al.}\cite{BraginskyDual} extended the theory to a dual recycling interferometer detector and showed that the multi-cavity coupling could reduce the effective linewidth to a sub-Hz range. If the high order cavity mode involved in parametric instability is resonant in both the arm cavities and the recycling cavity, extremely high three mode parametric gain could occur. Detailed analysis of a dual recycling interferometer with realistic test masses by  Gras \emph{et al.}\cite{GrasCQG1} showed that the highest gain could reach  $\sim1000$ corresponding to acoustic ring-up times $\sim$ seconds.

Recently modelling that takes into account large acoustic amplitudes, and using parameters close to those of Advanced LIGO (aLIGO) has shown that the growth of instability saturates. 
Danilishin \emph{et al.} showed that parametric instability is likely to grow on a time scale of minutes for realistic parameters \cite{sundae},\cite{Danilishin}.

Three mode parametric interactions are extremely sensitive to test mass mirror parameters. This extreme sensitivity was emphasised by Ju \emph{et al.}\cite{JuMonitor} who showed that mirror radius of curvature changes corresponding to wavefront deformations of $10^{-6}\lambda$ could easily be observed by monitoring three mode interactions in Advanced interferometers.

To date three-mode instability has been reported in one free space cavity experiment using a picogram membrane in a 10cm cavity \cite{sundae}, and in aLIGO \cite{matt}.

At the Gingin High Optical Power Facility \cite{gingin} a 74m optical cavity has been set up to be comparable to the conditions of Advanced interferometers. This paper is based on observations in this facility which, while demonstrating instability, has revealed a phenomenon that suppresses the exponential growth of instability at low amplitudes.

Previous modelling has ignored two real world aspects of practical suspended mass interferometers: a) that mirrors after coating have figure errors $\sim1nm$ RMS over the central diameter of 160mm, and b) that the laser spot position on the mirrors fluctuates due to residual low frequency seismic motion. The presence of figure errors means that the average radius of curvature of the region of the mirror intercepted by the laser beam depends on the beam location. This radius of curvature determines the transverse mode offset frequency. Because low frequency fluctuations of the spot position causes the laser spot to intercept different regions of the mirror surface, it follows that there will be dynamical modulation of the optical transverse mode offset frequency. The frequency modulation causes the parametric gain to be time dependent, and if the modulation amplitude exceeds the transverse mode optical linewidth, the gain can be strongly modulated.  This can create a situation where parametric instability does not have time to develop because it is only on-resonance intermittently, and for too short a time for instability to grow to problem levels.

In this paper we will show that the above phenomenon is likely to reduce the average parametric gain of the candidate modes most likely to become unstable, thereby significantly reducing the risk of instability. Results are confirmed by modelling and by measurements on a 74m optical cavity at Gingin. Recognition of this frequency modulation suppression mechanism also leads to methods by which suppression can be enhanced either by modulated thermal actuation or spot position dithering at frequencies below the gravitational wave sensitivity band.

In section \ref{secTheory} we summarise the theory of parametric instability and present modelling results showing how individual unstable modes can be suppressed by seismic induced frequency modulation. In section \ref{secFigError} we use aLIGO test mass mirror metrology data to estimate the frequency modulation expected for small spot position motions in aLIGO. In section \ref{secCavity} we present results obtained with the Gingin high optical power cavity: both the observation of parametric instability and the frequency modulation that greatly reduces the risk of instability. We discuss the results obtained, and their implications for aLIGO. We also present thermal actuation modelling results to estimate the suppression factors achievable.

\section{Theory of PI and effect of transverse mode frequency modulation}\label{secTheory}

Three mode opto-acoustic interactions occur when the frequency difference between an optical cavity pump mode at frequency $\omega_0$ and a transverse mode at frequency $\omega_1$ is appropriately tuned to the frequency of an acoustic mode at frequency $\Delta_m$. This three mode interaction resonance is defined by $\Delta_m=(\omega_0-\omega_1)-\omega_m=\Delta\omega-\omega_m=0$. The parametric gain $R$ characterises the ratio of acoustic energy input compared to mirror acoustic mode losses. If $R > 1$, the system is acoustically unstable, and the acoustic mode will grow exponentially until either non-linearities cause saturation \cite{Danilishin}, or else the cavity loses lock. In this paper we are concerned only with small amplitude excitation so can ignore non-linearities. The magnitude of R depends on cavity input power, on acoustic and optical mode losses, and on the spatial overlap between the relevant modes. For any pair of acoustic and optical modes, the gain R can be expressed as[3]

\begin{equation}
{\cal R}=\frac{P \Lambda\omega_1}
{M\omega_m L^2 \gamma_m \gamma_0\gamma_1}\frac{1}{1+(\Delta_m/\gamma_1)^2}, (\gamma_m\ll\gamma_1)
\label{eq:R_general}
\end{equation}

Here $P$ is the input power to the cavity, $\gamma_0$, $\gamma_1$, and $\gamma_m$ are the half-linewidth of the two optical modes and the acoustic mode of the test mass respectively, $M$ is the mass of the test mass, $L$ is the length of the cavity.  The $\Lambda$ is the overlap factor including the mass to the effective mass ratio as defined in \cite{Braginsky2001}.  The optical mode spacing $\Delta\omega$ is a function of the radius of curvature of the mirrors of the optical cavity and is given by

\begin{equation}
\Delta \omega=\frac{c}{L}(m+n)\cos^{-1} (\pm\sqrt{(1-\frac{L} {R_1})(1-\frac{L}{R_2})}),
\label{eq:f_gap}
\end{equation}
where $R_1$ and $R_2$ are the radii of curvature of the end mirrors of the cavity, m and n are integers describing the order of the optical mode. The ± sign depends on the cavity configuration. Equation (\ref{eq:f_gap}) assumes perfect spherical mirrors, but we will assume that in the case of figure errors the mode spacing is defined by the average radius of curvature at the laser spot position, averaged over the effective spot size.

Equation (\ref{eq:R_general}) considers only the Stokes process where parametric amplification or instability processes occur due to a single high order optical mode.  Here we want to focus particularly on the case where dynamic detuning causes $\Delta_m$ to be time dependent.  We consider the case of harmonic detuning given by

\begin{equation}
\Delta_m (t)= \Delta_{m0}cos\omega_d t,
\label{eq:detune1}
\end{equation}
where $\omega_d$ is a dynamic tuning frequency. In suspended mass interferometers the test mass-mirrors are supported by low frequency pendula which isolate against vibration. The test mass positions are controlled by feedback, but finite residual motion is inevitable because of the requirement that the test masses be inertial within the gravitational wave signal band.

Thus in practice, test masses can be expected to have significant motion at pendulum normal mode frequencies 0.1-1Hz. This gives rise to a modulation in the laser spot position. If the mirrors are imperfect, the mirror radius of curvature (averaged over the laser spot size) will vary smoothly with spot position. In this case, modulation in spot position can modulate the transverse mode offset frequency, thus causing time dependent detuning fluctuations.

Spot position motion will also modulate the modal overlap parameter. However for millimetre scale motions the overlap parameter modulation is small compared with the effect of detuning, and is ignored in the following analysis.

Assuming that $\Delta \omega_m$ changes according to Eq (\ref{eq:detune1}), the parametric gain is given by

\begin{equation}
R(t)= \frac{R_{max}}{1+(a\cos\omega_d t)^2},
\label{eq:detune2}
\end{equation}
where $a=\Delta_{m0}/\gamma_1$ is the normalised frequency detuning modulation amplitude. Equation \ref{eq:detune2} allows estimation of the effects of modulation on the growth of parametric instability. As discussed above,  modelling has shown that the characteristic ring up time scale for parametric instability in a detector similar to aLIGO is likely to be $\sim10^2$s \cite{Danilishin}. Since $\omega_d$ is fast compared with such ring up times, one would expect to observe modulated signal growth.

\begin{figure}[h]
\centering
\includegraphics[width=0.47\textwidth]{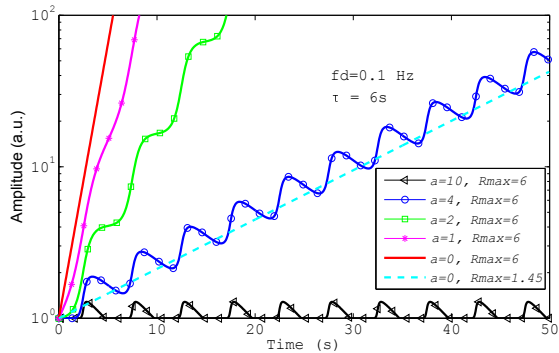}
\caption{Acoustic mode amplitude ring up curves for various detuning amplitudes.  Here we assume maximum gain $R_{max}$=6, an acoustic mode ring down time $\tau$=6 and a dynamic modulation frequency of 0.1 Hz.  For comparison, cases for on resonance ($a$=0) with $R_{max}$=6 and $R_{max}$=1.45 are also plotted.}
\label{figure:modulation}
\end{figure}

Figure \ref{figure:modulation} shows examples of possible acoustic mode ring up signatures. We assume parameters comparable to those of the experiment reported in this paper: $f_d=\omega_d/2\pi$ = 0.1Hz, $R_{max}=6$, and a normalised detuning amplitude $a$ = 1, 2, 4 and 10. It is sufficient to choose a typical acoustic mode decay time $\tau$ without need to specify acoustic mode frequency. We chose to use $\tau=6s$, corresponding to acoustic quality factor $Q_m=10^6$ and $4\times10^6$ for frequencies 50kHz and 200kHz respectively.  Results are compared with acoustic mode ring up curves in the absence of dynamic detuning ($a=0$) for $R=6$. In the case of $a=4$ the ring up slope is equivalent to that of a system with $a=0$ and $R=1.45$ as indicated in the figure. This represents a gain suppression factor $\sim4$.  Clearly in all cases, frequency modulation suppresses the effective parametric gain as determined by the average slope of the ring up curves.  For $a=10$ we see that instability has been replaced by a modulated acoustic mode amplitude which while not harmonic, is stable in time. The equivalent parametric gain $R_a$ in the presence of harmonic dynamic detuning is given by:

\begin{equation}
R_a= \frac{R_{max}}{\sqrt{1+a^2}},
\label{eq:Requ}
\end{equation}

\begin{figure}[t]
\centering
\includegraphics[width=0.4\textwidth]{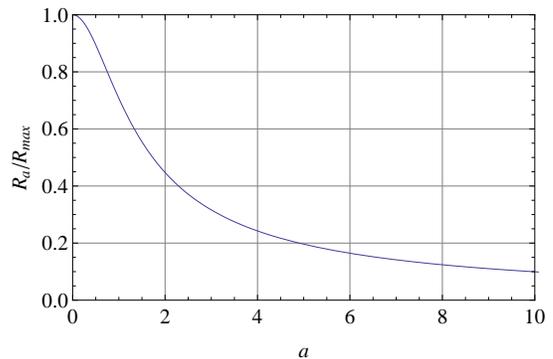}
\caption{Effective parametric gain suppression as a function of dynamic detuning amplitude $a$.}
\label{figure:Ra}
\end{figure}

The suppression of effective parametric gain as a function of modulation amplitude is shown in Fig. \ref{figure:Ra}.  Parametric gain can be suppressed by an order of magnitude for $a=10$.

\begin{figure}[t]
\centering
\includegraphics[width=0.45\textwidth]{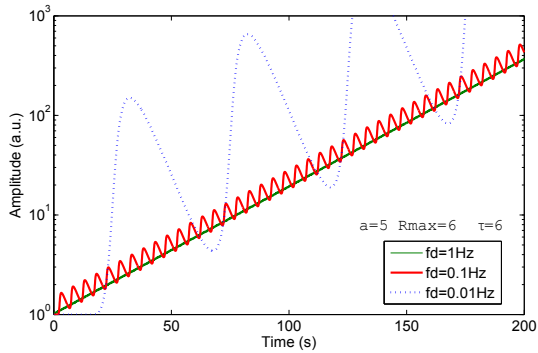}
\caption{As the dynamic detuning frequency increases the acoustic mode amplitude excursions are reduced but the effective parametric gain is unaltered. Here three detuning frequencies 1Hz, 0.1Hz and 0.01Hz are shown.  The detuning amplitude is fixed at $a=5$ ($R_{max}=6$ and $\tau=6$).   }
\label{figure:modu1}
\end{figure}

The mechanism discussed above occurs because the dynamic detuning modulation frequency is fast compared with the acoustic mode ring up time scale. The observed acoustic mode amplitude modulation occurs at double the dynamic detuning frequency $\omega_d$. While the effective parametric gain is independent of $\omega_d$, the peak to peak acoustic mode amplitude within one cycle is inversely dependent on $\omega_d$. Figure \ref{figure:modu1} shows some examples for three different dynamic detuning frequencies. The amplitude modulation waveform is highly non-linear since it is due to a Lorentzian modulation acting on the exponent.

\begin{figure}[b]
\centering
\includegraphics[width=0.45\textwidth]{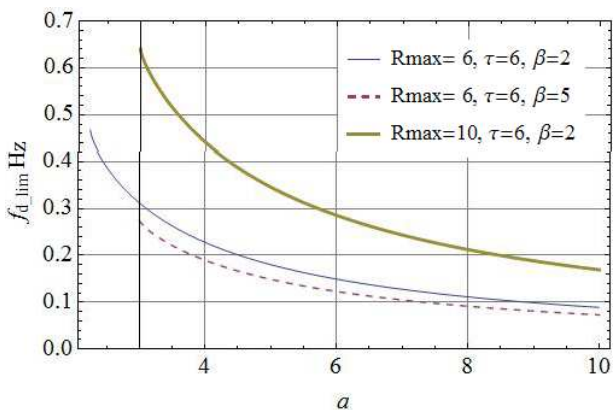}
\caption{Dynamic detuning frequency limit to prevent amplitude excursions exceeding a predetermined value $\beta$. For example, for $R_{max}=10$ and $\tau=6s$, with an amplitude growth requirement of $\beta=2$, then the minimum dynamic detuning frequencies is limited to between 0.6 Hz and 0.1 Hz for detuning amplitude $a$ in the range of 2 and 16.}
\label{figure:flim}
\end{figure}

It can be seen that if the detuning frequency is too slow, the acoustic mode amplitude can grow to a very large value within half a detuning period.  It is possible to define a lower limit for the dynamic detuning frequency $f_{d\_lim}$ to prevent the acoustic mode amplitude excursion from exceeding $\beta$ times its original value within one cycle. Figure \ref{figure:flim} shows three curves showing the lower limit of the dynamic detuning frequency as a function of detuning amplitude $a$, for two values of $R_{max}$ and two values of the acoustic amplitude excursion limit $\beta$. For example if  $R_{max}=10$, and $\tau=6s$, and a requirement $\beta=2$, then the dynamic detuning frequency is limited within the range 0.1Hz - 0.6Hz assuming detuning modulation amplitudes $a$ is between 2 and 16. We see that larger detuning frequencies or larger detuning amplitudes both act to reduce amplitude excursions. This defines the parameter space for suppressing parametric instability by the dynamic detuning mechanism.
In the next sections we will see that the above mechanism can occur naturally as a result of residual motion in the presence of mirror figure errors, which we consider in the context of aLIGO and a 74m cavity at Gingin.

\section{Frequency modulation by mirror figure errors\label{secFigError}}

The test mass mirrors in aLIGO have radii of curvature of $\sim 2000m$. Figure errors mean that the effective radius of curvature depends on the spot position. For example, using equation \ref{eq:f_gap}, with aLIGO arm cavities, if the ETM RoC changes 1m from nominal value of 2242m (corresponding to a sagitta change within a beam diameter $\sim$ 0.3nm), the TEM10 cavity mode frequency will change $\sim$ 13Hz.

The residual motion of the test masses in interferometer arm cavities therefore cause high order mode frequency modulation. Residual angular motion creates beam residual motion on the test mass surface ~ millimetres \cite{Fritschel}. Depending on the test mass figure errors, this residual motion causes dynamic detuning of the cavity high order mode frequency at the frequencies of test mass pitch and yaw motion.
\begin{figure}[h]
\centering
\includegraphics[width=0.4\textwidth]{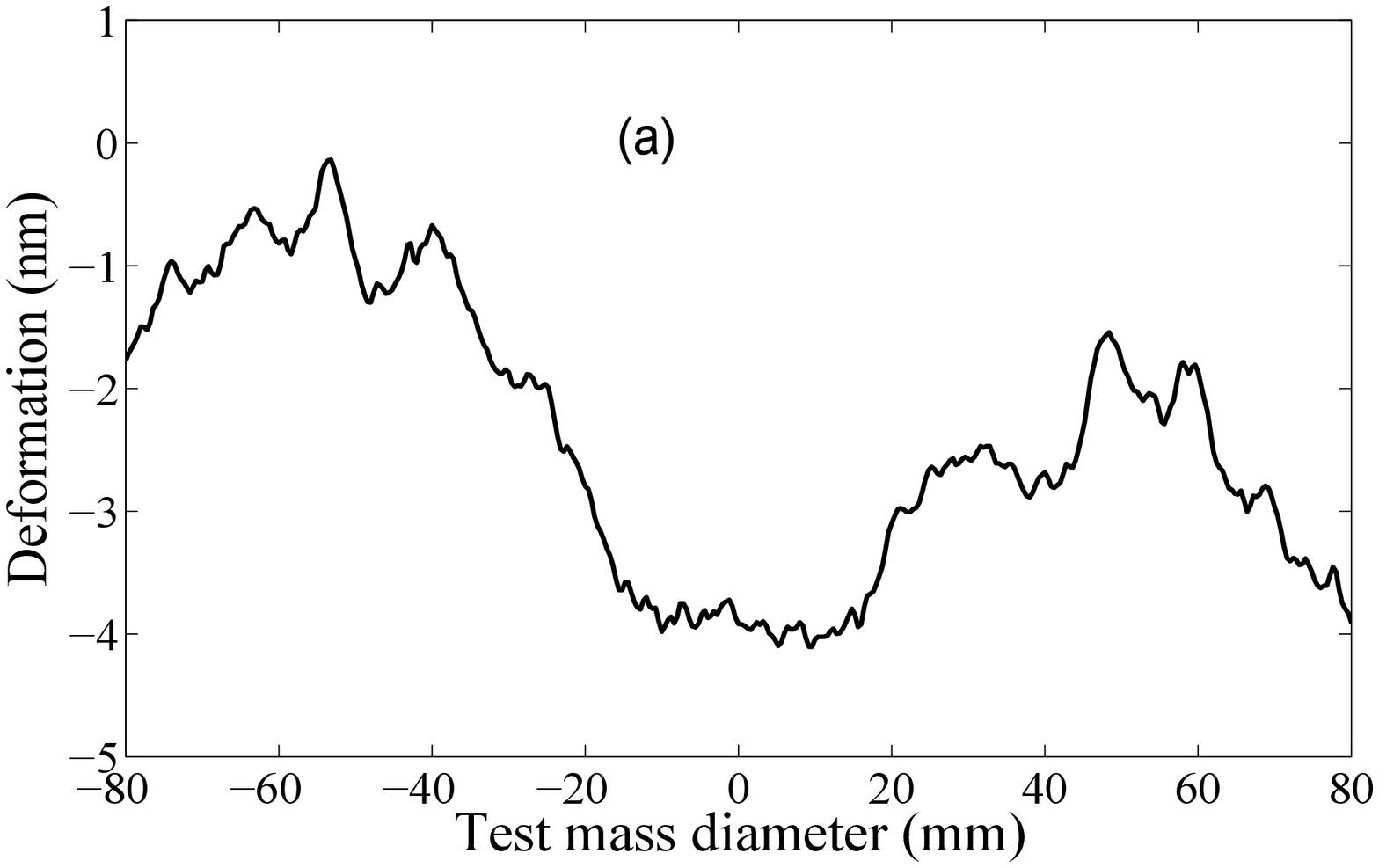}
\includegraphics[width=0.4\textwidth]{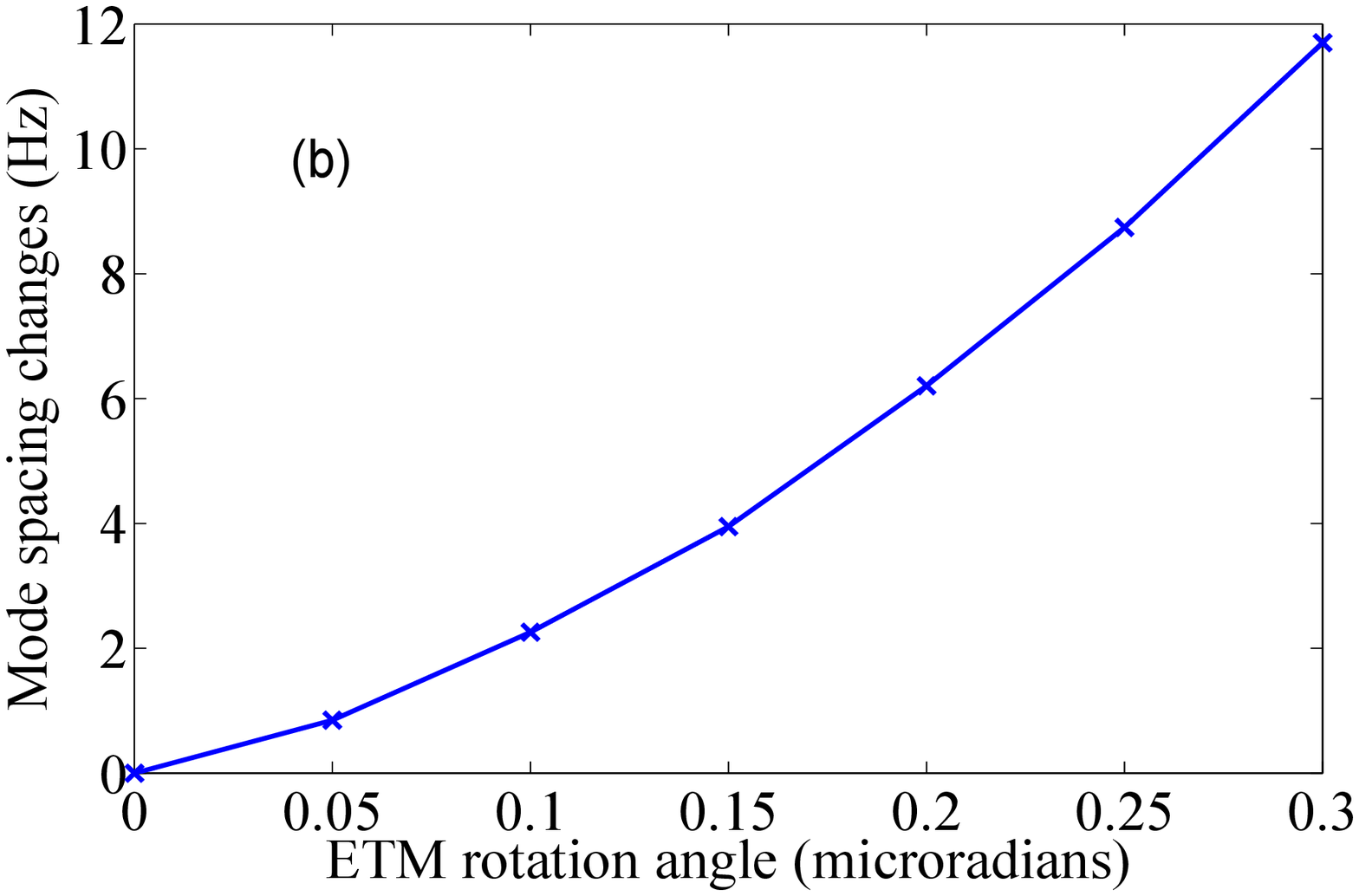}
\caption{a) Typical aLIGO test mass figure errors (compared to a perfect sphere of RoC $\sim 2242m$) showing deformations across a mirror diameter. b) FFT code model for frequency offset as a function of test mass angular motion.}
\label{figure:profile12}
\end{figure}

To estimate the aLIGO arm cavity high order mode frequency changes as a function of the residual test mass angular motion, we used measured test mass surface data \cite{ITMETM} in an interferometer simulation codes (OSCAR \cite{Oscar} and FOPG\cite{FOGP}) to simulate cavity transverse mode detuning. For the simulation we fixed the ITM and modelled ETM misalignment at various angles from zero to 0.25 micro-radians. Figure \ref{figure:profile12}(a) shows an example of the input data in the form of a cross section across a test mass diameter showing how the figure errors increase with radius. Figure \ref{figure:profile12}(b) shows the calculated data for real 2D surface profiles.

Figure \ref{figure:profile12} (b) shows that the cavity mode spacing increases roughly quadratically with ETM misalignment angle. Note that 0.1 microradian corresponding to $\sim$2mm beam position displacement on the test mass in a typical aLIGO arm cavity. We extended the simulation to the test mass rotation corresponding to a spot displacement $\sim$6 mm. We note that in reality both test masses move independently of each other, thereby creating somewhat larger detuning amplitudes.

\begin{figure}[h]
\centering
\includegraphics[width=0.4\textwidth]{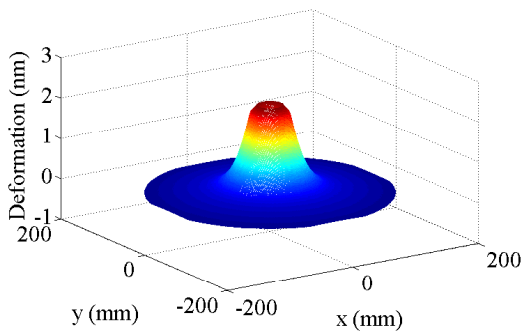}
\caption{The maximum thermal deformation when 0.1 Hz sinusoid heating power of amplitude 2W applied on the front surface of the test mass}
\label{figure:profile3}
\end{figure}

Cavity high order mode frequency modulation can also be artificially created by applying modulated heating to the test mass. We used ANSYS FEM simulation software package to simulate the transient thermal deformation of the test mass surface under sinusoid heating power. Figure \ref{figure:profile3} shows the maximum thermal deformation when a 0.1 Hz modulated heating beam of 50mm radius with 2W peak-to-peak power amplitude is applied to the test mass front surface. This deformation corresponds to a cavity mode spacing frequency change of $\sim$40Hz in aLIGO arm cavity simulated using FFT code\cite{Oscar}.

The above results indicate that passive detuning frequency modulation in aLIGO would be expected to be $\sim$few Hz for the TEM10 transverse mode, which is much smaller than the arm cavity linewidth and does not have significant effect on parametric gain. However this could be increased to $\sim$40Hz modulation using CO2 laser heating. It is important to note that the highest predicted parametric gains in aLIGO are for modes up to 4th order \cite{Evans}. By equation \ref{eq:f_gap} detuning scales with mode order. 
Thus the above estimates would correspond to at least 4 times larger modulation ($\sim$160Hz) for 4th order instabilities.
The aLIGO arm cavity half-linewidth is $\sim$40Hz. Thus the parametric gain associated with arm cavity optical modes would be suppressed by factors $\sim$few, and for low order cavity modes the above modulations could be negligible.
Reference \cite{Evans} shows that the highest parametric gain instabilities are associated with modes that are resonant in the power recycling cavity. For these modes, the coupled cavity linewidth is about 0.3 Hz \cite{GrasCQG1}, the normalised dynamic detuning amplitude can exceed 50 times of the coupled cavity linewidth. Thus it is most likely that intrinsic passive detuning in aLIGO will lead to a parametric gain suppression factor $>100$ for those modes resonant inside the recycling cavity. A detailed simulation to explore how the cavity high order mode frequency modulation affects the broad spectrum of parametric instability in aLIGO is beyond the scope of this paper. However the experimental observation of parametric instability presented in the next section largely confirms the above theory.

\section{High Optical Power Cavity Observations}\label{secCavity}

\begin{figure}[t]
\centering
\includegraphics[width=0.4\textwidth]{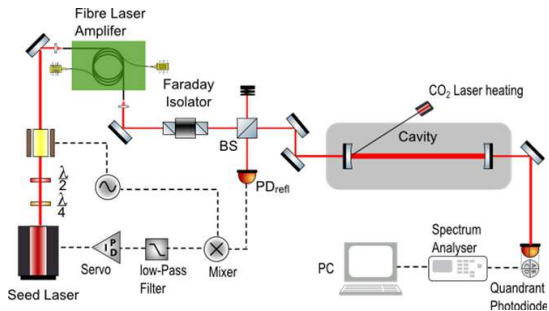}
\caption{ Schematic diagram of the experimental setup: The laser light from a seed laser is amplified by a 50 W fibre laser amplifier. The high optical power laser beam is injected into the 74m long optical cavity. The seed laser is frequency locked to the long cavity using PDH locking. The cavity transmitted beam is detected by a quadrant photodiode (QPD). The differential signal from the QPD measures the beating between the cavity fundamental mode and the first order mode.}
\label{figure:setup}
\end{figure}

We studied three mode parametric instabilities at the Gingin high optical power facility. The experimental setup is shown in Figure \ref{figure:setup}. A 74m long optical cavity with fused silica test masses is suspended from high performance vibration isolators \cite{Barriga, Dumas} by a modular 4-wire test mass suspension system developed at UWA. The test masses are installed in two large vacuum chambers connected by a 400mm vacuum pipe. The system was assembled in clean room conditions and uses a hydrocarbon free vacuum system to enable high optical power densities to be achieved.

Both test masses are 50mm in diameter and 50mm thick, with mass $\sim$0.8kg. The nominal $RoC$ of the two test masses are 37.5m and 37.4m. The test masses have a very sparse mode spectrum compared to aLIGO test masses, so that three mode parametric interactions need to be tuned to specific candidate acoustic modes. This is achieved by using a power stabilised CO2 laser to thermally tune the ITM RoC to create three mode tuning for the specific candidate acoustic modes \cite{Susmithan1}. The dominant test mass residual angular motions are at frequencies of 0.15 Hz.

The measured cavity finess is 14500$\pm$300. The light source is a 50W fibre laser amplifier fed by a 400mW Nd:YAG NPRO seed laser. The seed laser is frequency locked to the long cavity using PDH locking \cite{PDH}. The cavity transmission is detected by a quadrant photodiode (QPD). The differential output of the QPD measures the beating between the cavity fundamental mode and the first order mode while the sum of the QPD output measures the total cavity transmitted power. A spectrum analyser (Agilent 89410A) and a PC are used to analyse and to record the signal.

\begin{figure}[t]
\centering
\includegraphics[width=0.25\textwidth]{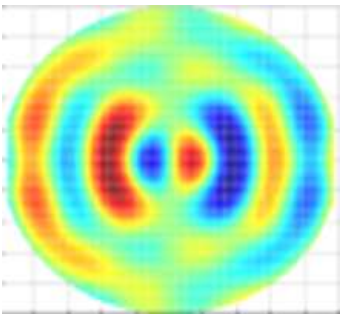}
\caption{The test mode acoustic mode amplitude distribution on the surface.}
\label{figure:map}
\end{figure}

Using the ANSYS software package, we first analysed the test mass acoustic mode structure and frequencies. Based on the simulation we then identified one particular acoustic mode that has good overlap with the cavity first order mode, and minimum vibration amplitude at the suspension point to minimise the mechanical loss introduced by the suspension.  Our target mode, with simulation frequency 150.49 kHz is in the range for easy $CO_2$ laser thermal tuning. The mode amplitude distribution on the test mass surface is shown in Fig. \ref{figure:map}.  The overlap factor taking into account the mode effective mass is $\sim$16. The measured mode frequency is $\sim$150.2 kHz (depending on the temperature.) The measured mechanical Q-factor using the ring-down method is $\sim3.4 \times 10^6$.

Three mode interaction conditions are achieved by tuning the TEM00 and TEM10 mode spacing close to 150.2kHz using CO2 laser thermal tuning. Measurement of the tuning is relatively easy because residual laser beam-jitter noise gives rise to a small amount of TEM10 mode power inside the cavity which beats with the TEM00 at the QPD, allowing the TEM10 offset frequency to be monitored as a beat note. This provides a means for monitoring the mode spacing by measuring the cavity transmitted power on the QPD where the two modes are mixed.

\begin{figure}[t]
\centering
\includegraphics[width=0.4\textwidth]{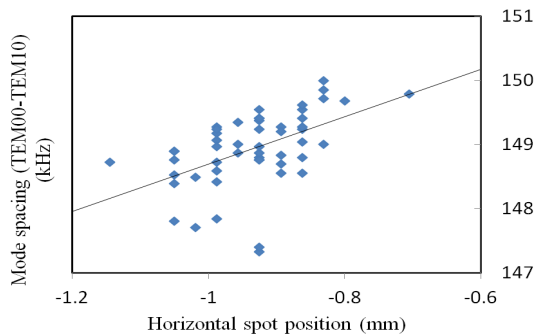}
\caption{The correlation between horizontal spot position of the ITM alone and the transverse mode frequency indicates that the frequency detuning is caused by the spot position change. The solid line is the linear fit to the measurement data. The spread of the data is due to the fact that the laser spot position is also change in ETM.}
\label{figure:cavitymodes}
\end{figure}

The mode spacing was observed to fluctuate with a typical peak to peak amplitude $\sim$ few kHz. To confirm that these fluctuations were associated with beam spot position on the test masses, we recorded the cavity mode spacing and the beam position simultaneously for the ITM. Figure \ref{figure:cavitymodes} shows the mode spacing as a function of the beam position on the ITM in horizontal direction.  The beam position was determined by recording the video of the CCD camera and then analysed by referencing it to the test mass diameter.  In the horizontal direction, there is a linear correlation between increased mode spacing with increased beam position. The solid line in figure \ref{figure:cavitymodes} is a linear least squares fitting to the measurement data. The relative large scatter is due to the fact that we recorded only the ITM beam position while the ETM beam position is also not stable. The effect is more difficult to measure in the other axis because the suspensions introduce much smaller vertical beam position fluctuations. However the single axis correlation is sufficient to confirm our conjecture that mirror figure errors translate into dynamic detuning.

We do not have precise metrology of our test mass mirror profiles. However the observed fluctuations are consistent with the mirror figure error specification of 1nm. It is interesting to note that in principle simultaneous measurement of spot position on both test masses and transverse mode frequency offset could be used to allow precise metrology of both test masses.

When the cavity is correctly tuned the three mode interaction occurs, and the signal at the QPD becomes dominated by the beating between TEM00 and TEM10 modes at the acoustic mode frequency. The signal is proportional to the acoustic mode amplitude, the TEM00 mode power and the TEM10 mode detuning. The signal is normally most easily observed by mixing the acoustic frequency with a local oscillator, combined with a low pass filter, so as to reduce the signal frequency to $<10$Hz.

As discussed above residual motion causes cavity detuning. The residual motion amplitude depends on  environmental noise, which excites the suspension normal modes. Most of the time we observe dynamic detuning with a frequency amplitude of 1-5 kHz. Even under these circumstances the acoustic mode signal at frequency $\sim$150.2 kHz can normally be clearly observed.

Wind forces on the laboratory, microseismic activity and human activity all contribute to degrading the residual motion. During quiet times the residual motion is reduced and for times of $\sim$30 seconds the detuning amplitudes can be less than a few cavity linewidths. In these short periods of time conditions are suitable for observing three mode parametric instability.

\begin{figure}[t]
\centering
\includegraphics[width=0.48\textwidth]{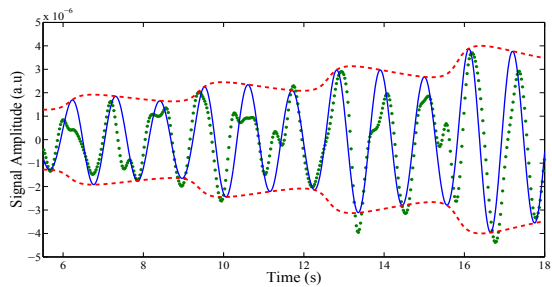}
\caption{The QPD differential output signal at test mass acoustic mode frequency (150.28 kHz). The signal was down-converted to $\sim$0.91 Hz by mixing with an local oscillator signal. The solid line is a fitting curve of 0.91Hz with parametric gain $R_{max}=6$ and detuning amplitude a=2. The growing signal envelope (dashed line) is consistent with suspension modulation at 0.15Hz. The effective parametric gain is $\sim$1.45.}
\label{figure:ringup}
\end{figure}

To observe the signature of parametric instability we increased the cavity circulating power to $\sim30$ kW. For periods of time $\sim$10 to 30 seconds, when the dynamic detuning is low, the acoustic signal can be observed ringing up with time, as shown in figure \ref{figure:ringup}. In this case the acoustic signal frequency was down converted to 0.91 Hz as discussed above.

Observations under best tuned quiet conditions show the acoustic signal growing for times $\sim$ 14 seconds. The amplitude growth is modulated but more complex than the single modulation frequency model used in section \ref{secTheory}, due to the presence of several low frequency modulation frequencies associated with the angular motion both test masses. Beating also occurs, due to the fact that the two test masses have closely spaced suspension normal modes. The beating causes the detuning amplitude to vary periodically over time scales $\sim$ 30 seconds. The effective parametric gain based on observed ring-ups during times of minimum detuning amplitude such as shown in Figure \ref{figure:ringup} is R $\sim$1.45.

In Figure \ref{figure:ringup} we have fitted to a double frequency of a single 0.15Hz suspension mode to model the dynamic detuning. This gives a modest fit to the data but a complete fit is not possible due to the stochastic nature of the seismic excitation of the normal modes.

 \section{Conclusions}\label{secConc}

 We have created conditions in which three mode parametric instability can occur in a suspended high power optical cavity designed to mimic conditions comparable to those expected in advanced gravitational wave detectors. We have observed time dependent growth of a 150.2kHz acoustic mode, consistent with a new model of parametric instability for suspended mass optical cavities. The gain in the parametric instability regime is lower than previously expected, and modulated by low frequency residual motion. Results are consistent with a new model for the build up of instability in which transverse mode frequency fluctuations act to reduce the parametric instability power build up through dynamic detuning which itself is caused by residual motion in the presence of nm-level mirror figure errors. Data on aLIGO optical cavities indicate that the same phenomenon will act to reduce the risk of parametric instability for the highest parametric gain modes. Mirror imperfections have beneficial effects in this regard. Results also point to simple methods for reducing parametric gain by thermal modulation of test masses or by low frequency dithering of the test masses. Further studies on full scale detectors to quantify the dynamic detuning and linewidths of transverse modes are needed to quantify these effects.

{\bf Acknowledgements}

We wish to thank the Gingin Advisory Committee of the LIGO Scientific Collaboration and the LIGO Scientific Collaboration Optics Working Group for encouragement. Thanks to our collaborators Jesper Munch, Peter Veitch and David Hosken for useful advice. We wish especially to thank  Slawek Gras for his careful review of the manuscript and the LIGO MIT group for their encouragement. This research was supported by the Australian Research Council.

\end{document}